\documentstyle[aps,prl,twocolumn,floats,epsf]{revtex}
\begin{document}
\draft

\title{Role of Metastable States in Phase Ordering Dynamics}
\author{R. M. L. Evans, W. C. K. Poon and M. E. Cates}
\address{Department of Physics and Astronomy, The University of Edinburgh,
Edinburgh EH9 3JZ, U.K.}

\date{January 20, 1997}
\maketitle

\begin{abstract}
We show that the rate of separation of two phases of
different densities (e.g.\ gas and solid) can be radically altered by the
presence of a {\em metastable} intermediate phase (e.g.\ liquid).
Within a Cahn-Hilliard theory we study the growth in one dimension
of a solid droplet from a supersaturated gas. A moving interface between
solid and gas phases (say) can, for sufficient (transient) supersaturation,
unbind into two interfaces separated by a slab of metastable liquid phase. We
investigate the criteria for unbinding, and show that it may
strongly impede the growth of the solid phase.
\end{abstract}


  The influence of metastable states on the dynamics of phase transformations
is an important issue in materials physics. Their role has long been
acknowledged in metallurgy, where the rate of phase transformation is often
limited by conduction of (latent) heat \cite{Cahn68,DeHoff93}. Indeed,
Ostwald's empirical `rule of stages' \cite{Ostwald} asserts that the
transformation from one stable phase to another proceeds via all metastable
intermediates in turn. In complex fluid systems such as colloidal suspensions,
particle diffusion, not heat diffusion, is often rate-limiting \cite{caveats}.
Here too, a strong influence of metastable phases on ordering dynamics has
been suggested, e.g.\ in the phase ordering kinetics of polymer liquid
crystals \cite{Heberer95}, random-coil polymers \cite{Berghmans95}, proteins
\cite{PoonPRE} and colloid-polymer mixtures \cite{Poon95,vanBruggen96}.
In this Letter, we propose a simple model to account for the role of
metastable states in phase ordering limited by particle diffusion (described by
a conserved order parameter; for the nonconserved case, e.g.\ nematic order,
see Refs.\ \cite{Bechhoefer91}).

\begin{figure}[h]
  \epsfxsize=8cm
  \begin{center}
  \leavevmode\epsffile{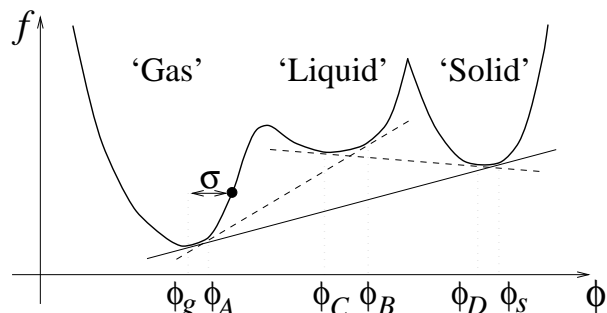}
  \caption{Bulk free energy density $f$ versus concentration $\phi$, with
   double tangents for equilibrium binodals (solid line), and metastable
   binodals (dashed lines). The supersaturation $\sigma$ for an initial
   homogeneous state ($\bullet$) is defined as shown.}
  \label{feexamp}
  \end{center}
  \vspace{-12pt}
\end{figure}

  To establish ideas,
consider first a homogeneous fluid of a simple substance which
is quenched to a temperature below its triple point. It will separate into two
coexisting phases, the gas and the solid, whose densities are given by the
construction in Fig.\ \ref{feexamp}. However, in this
temperature range, the low density (fluid) branch of the free energy has an
additional minimum, representing the metastable liquid phase, whose presence
can interfere with the phase separation process. Free energy curves of
precisely this form can be realized in several complex fluid systems, notably
mixtures of spherical colloids and much smaller polymers \cite{cpfoot}.
Experiments \cite{Poon95} show that, when such a mixture is prepared with a
composition at which a metastable minimum is present \cite{cpfoot} it does not
separate quickly into dilute colloidal fluid and dense colloidal crystal
(the predicted equilibrium phases). Instead, after an initial latency period
(consistent with nucleation) dense but non-crystalline droplets separate out,
forming an amorphous sediment which begins to crystallize only on a much longer
time-scale \cite{nematic}.

  We can explain these qualitatively robust results within a simple,
one-dimensional continuum model if we take due account of the
metastable `liquid' minimum in the free energy curve. (We now use `gas, liquid,
solid' as {\em generic} labels for three phases of increasing densities
\cite{nematic}.) Details of our results will be given elsewhere
\cite{Evans97a,Evans97b}.

\begin{figure}
  \epsfxsize=8cm
  \begin{center}
  \leavevmode\epsffile{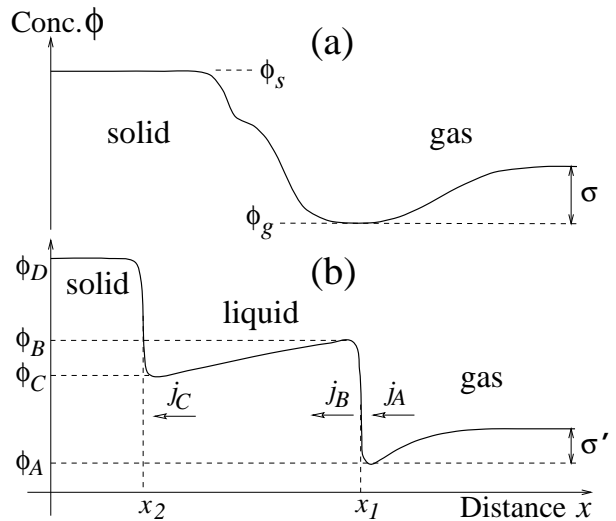}
  \caption{(a) A solid-gas interface. Note the inflection near the density of
   the `liquid minimum' in $f(\phi)$. (b) A split interface: the inflection has
   become a slab. Solid, metastable liquid and gas regions are separated by
   `curvature-unbound' interfaces at $x=x_{1}(t)$ and $x_{2}(t)$, between
   binodal densities $(\phi_{A},\phi_{B})$ and $(\phi_{C},\phi_{D})$
   respectively (see Fig.\ \ref{feexamp}). Fluxes $j_{A}$, $j_{B}$, $j_{C}$ flow
   in and out of the interfaces. In the gas phase, the supersaturation
   {\em relative to the metastable binodal} $\phi_{A}$, is $\sigma'$.}
  \label{interfaces}
  \end{center}
  \vspace{-16pt}
\end{figure}

  Consider a system with a conserved order parameter (a concentration or density
$\phi$ \cite{caveats}) at equilibrium. At coexistence, there is an
interface between two phases (`gas' and `solid') whose densities ($\phi_{g}$
and $\phi_{s}$) are found from the bulk free energy density $f(\phi)$, as in
Fig.\ \ref{feexamp}, by the double-tangent construction \cite{DeHoff93}, which
ensures equality of the chemical potential $\mu$ and the (osmotic)
pressure $P$ in the two phases. ($\mu$ is the slope of the tangent and
$P$ its negative intercept.) In minimizing the free energy, we seek the lowest
possible double tangent, and may safely ignore local minima such as that
labeled `liquid' in Fig.\ \ref{feexamp}.
Across the interface, $\phi$ changes smoothly (as in Fig.\ \ref{interfaces}a),
passing through values for which the $\mu$ and $P$ calculated from $f(\phi)$
are ``wrong".  However, the curvature of the concentration, $\nabla^2\phi$,
contributes to $\mu$ and $P$ so that both are constant for the smooth interface
which exists in the thermodynamic ground state.

  Similar interfaces form in the early stages of phase separation, between
regions of different $\phi$ \cite{Langer92}. In the so-called
``growth stages" $\phi(\vec{x},t)$ evolves primarily by motion of these
interfaces \cite{Bray94}. In studying this motion, metastable minima are
typically ignored but, as we now show, they can play a crucial role.

  For simplicity we consider only one spatial dimension (perpendicular
to an interface) and study the evolution of the interfacial density profile.
The Cahn-Hilliard equation \cite{Cahn58} for this time evolution is
\begin{equation}
\label{CH}
  \frac{\partial \phi}{\partial t} = \frac{\partial}{\partial x} \left\{ \Gamma
  \frac{\partial}{\partial x}\left( \frac{df(\phi)}{d\phi}
  - K \frac{\partial^2\phi}{\partial x^2} \right)\right\}
\end{equation}
where the mobility $\Gamma$ and curvature constant $K$ are
phenomenological parameters; we are interested in $f(\phi)$ of the form shown
in Fig.\ \ref{feexamp}. Eq.\ (\ref{CH}) is derived by equating
$\partial\phi/\partial t$ to
the divergence of a current $J=-\Gamma\nabla\mu$ induced by a chemical potential
gradient. The chemical potential is defined in terms of a functional derivative,
$\mu=\delta F[\phi]/\delta\phi$, where $F[\phi]$ is the integral of the local
free energy density, $f(\phi)+K(\nabla\phi)^2/2$.

  Consider first an ordinary, smooth solid-gas interface
as shown in Fig.\ \ref{interfaces}a. If the solid region is to grow, the
interface must move to the right, and this requires a chemical potential
gradient in the supersaturated gas phase. Steady-state
solutions of Eq.\ (\ref{CH}) (a uniformly moving interface) are thus unphysical
{\em in far field}, since a non-zero gradient would have to extend to infinity.
However, such solutions do shed light on the {\em local} dynamics of the
interface since, during the growth stage
\cite{Tokuyama93}, local re-arrangements are much quicker than changes in the
far diffusion field driving the interfacial motion. Hence the interface
experiences a quasi-constant flux from the supersaturated region; its local
dynamics are quasi-steady-state \cite{Evans97a}.
Steady-state solutions of Eq.\ (\ref{CH}) are calculable to
arbitrary accuracy for any form of $f(\phi)$ and $\Gamma(\phi)$, and exactly
for piecewise-quadratic $f(\phi)$ and piecewise constant $\Gamma$
\cite{Evans97a}. (Constant $K$ is assumed.) In the latter case, linear
stability of an ordinary moving interface (Fig.\ \ref{interfaces}a) can be
established rigorously \cite{Evans97a}. Despite this, we
now explore an alternative, `split' mode of interfacial motion.

  Since the
double-tangent construction identifies states of equal chemical potential and
pressure, an interface may form between points on the metastable binodal, such
as $(\phi_{A},\phi_{B})$ in Fig.\ \ref{feexamp}. So, a split profile such as
that shown in Fig.\ \ref{interfaces}b, with well-separated gas-liquid (g/l)
and liquid-solid (l/s) parts, has only gentle gradients of $\mu$, and is
therefore long-lived. Physically,  a moving gas-solid (g/s) interface
(Fig.\ \ref{interfaces}a) is {\em locally} stable against this splitting
\cite{Evans97a},
because, if the flux of material from the right onto the g/s interface is
increased (so as to advance the base more than the top,  splitting the
interface) the curvature, $\nabla^2\phi$, is perturbed, setting up currents
tending to restore the profile. The g/l and l/s interfaces can be
described as {\em curvature-bound}. This {\em negative} feedback mechanism
does not operate for the well-split interface of Fig.\ \ref{interfaces}b. In the
metastable liquid region, curvature is small, and the Cahn-Hilliard equation
reduces to the diffusion equation \cite{Langer92}. Material flows to the l/s
interface from the g/l interface down the intervening concentration gradient,
{\em whose sign follows directly from the metastability of the liquid},
as can be seen
from the double tangent constructions in Fig.\ \ref{feexamp}. If the flux of
material onto the g/l interface is sufficient to increase its separation from
the solid droplet, the concentration gradient in the middle region, and hence
the restoring flux, will be reduced. So a {\em positive} feedback mechanism
operates; the split interface is then {\em
curvature-unbound}.

\begin{figure}
  \epsfxsize=8cm
  \begin{center}
  \leavevmode\epsffile{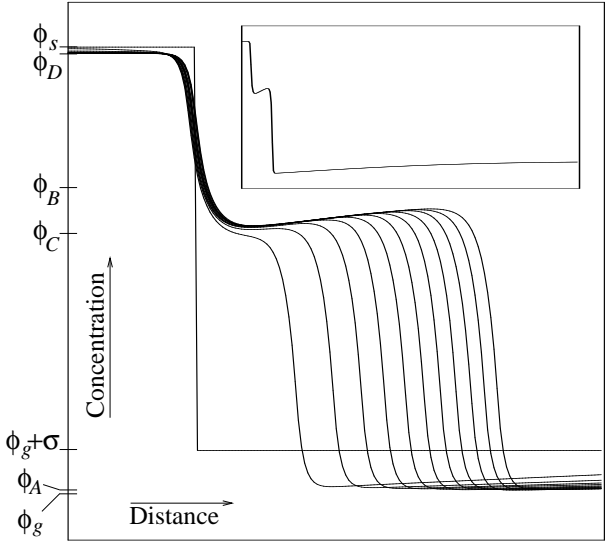}
  \caption{Superimposed snap-shots at equal time intervals, from numerical
   solution of Eq.\ (\protect\ref{CH}), using $f(\phi)$ as sketched in
   Fig.\ \protect\ref{feexamp}. Initial condition is a sharp step between
   $\phi_{s}$ and $\phi_{g}+\sigma$. A region of metastable liquid forms
   as the two parts of the interface ``curvature-unbind". Inset: Wide-frame
   view of the last time-slice.}
  \label{timeseq}
  \end{center}
  \vspace{-16pt}
\end{figure}

  We have observed curvature-unbinding in numerical solutions of
Eq.\ (\ref{CH}), using $f(\phi)$ of the form sketched in Fig.\ \ref{feexamp}. A
time-sequence is shown in Fig.\ \ref{timeseq}; the initial
condition is a sharp step between a solid at density $\phi_{s}$ and a
supersaturated state ($\phi=\phi_{g}+\sigma$) within the gas (rather than
liquid) well of the free energy curve. Below some critical supersaturation
(discussed below), such curvature-unbinding does
not occur, and the interface propagates in the conventional (unsplit) mode.

  We now analyse the dynamics of a pair of curvature-unbound interfaces.
On scales much larger than the characteristic interfacial width $\sqrt{K/f''}$,
the g/l and l/s interfaces can be represented by sharp steps at
positions $x_{1}(t)$ and $x_{2}(t)$ respectively (Fig.\ \ref{interfaces}b).
Concentrations at the interfaces will be fixed at the metastable binodal values
(Fig.\ \ref{feexamp}), so long as interfacial velocities are small
($\dot{x_{1}},\dot{x_{2}}\ll |f''|^{3/2}K^{-1/2}\,\Gamma$).
In the gas and liquid regions, evolution is purely diffusive, with diffusion
constants $D_{g}$ and $D_{l}$ respectively; $D$ is set to zero in the solid
phase. Conservation of material requires
$(\phi_{B}-\phi_{A})\dot{x}_{1}=j_{A}-j_{B}$ and
$(\phi_{D}-\phi_{C})\dot{x}_{2}=j_{C}$, where the fluxes $j_{A}$, $j_{B}$ and
$j_{C}$ are defined in Fig.\ \ref{interfaces}b.

  The control parameter is now the supersaturation  {\em relative to the
metastable binodal} $\sigma'\equiv\sigma-(\phi_{A}-\phi_{g})$ (see Fig.\
\ref{feexamp}). Two
further approximations are made, whose adequacy has been checked by numerical
solution of the full equations \cite{Evans97b}. The first is that the diffusive
field in the gas phase is unperturbed by the motion of the steps, leading
\cite{Evans97b} to the equation $j_{A}\approx\sigma'\sqrt{D_{g}/\pi t}$. The
second is that the gradient in the liquid region is approximately uniform,
leading to $j_{B}\approx j_{C}\approx D_{l}\,(\phi_{B}-\phi_{C})/(x_{1}-x_{2})$.
These yield tractable equations of motion for the size of the
metastable liquid region:
\begin{equation}
\label{metastable}
  \frac{d(x_{1}-x_{2})}{d(t/\tau)} = (t/\tau)^{-\frac{1}{2}} -
  \frac{(\sigma_{c}/\sigma')^2}{2(x_{1}-x_{2})}
\end{equation}
and for the size of the solid droplet:
\begin{equation}
\label{x2}
  x_{2}(t) = \left( \frac{\phi_{B}-\phi_{C}}{\phi_{D}-\phi_{C}}\right) D_{l}
  \int_{0}^{t}\! \frac{dt'}{x_{1}(t')-x_{2}(t')}.
\end{equation}
Here $\tau \equiv {\pi}({\phi_{B}-\phi_{A}})^2/(D_g{\sigma'}^2)$ and
$\sigma_{c} \equiv (\phi_{B}-\phi_{A})\times$
$\left\{ 2\pi(\phi_{B}-\phi_{C}) \left[({\phi_{B}-\phi_{A}})^{-1}
+({\phi_{D}-\phi_{C}})^{-1}\right]D_l/D_g\right\}^{1/2}$.

\begin{figure}
  \epsfxsize=8cm
  \begin{center}
  \leavevmode\epsffile{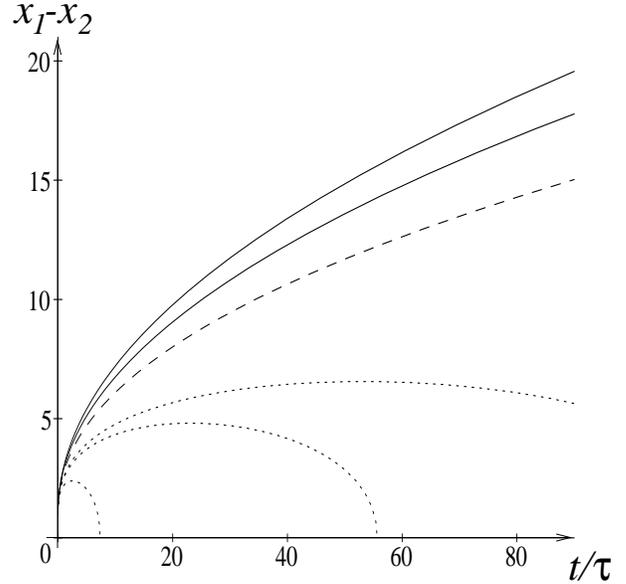}
  \caption{Size of the metastable region $x_{1}-x_{2}$ against
   time $t$. Solutions of finite lifetime (dotted lines) for
   $\sigma'/\sigma_{c}=0.58$,
   $0.70$ and $0.75$. The marginal solution $\sigma'=\sigma_{c}$ (dashed
   line). Diffusively unbound solutions (solid lines) for
   $\sigma'/\sigma_{c}=1.1$ and
   $3.2$. In each case $x_{1}-x_{2}=1$ at $t=0$.}
  \label{trajectories}
  \end{center}
  \vspace{-16pt}
\end{figure}

  For $0\leq\sigma'<\sigma_{c}$, solutions of Eq.\ (\ref{metastable}) collapse
to zero size in a finite time, as might be expected of a metastable
phase. We describe these interfaces as {\em diffusively bound}. (This is a much
``looser" binding than the curvature binding discussed above.) After
collapse of the metastable slab, the interfaces will curvature-recombine, and
the solid will resume the ordinary (unsplit) mode of growth. In contrast,
for $\sigma'>\sigma_{c}$ solutions of Eq.\ (\ref{metastable}) have open
trajectories. The interfaces are {\em diffusively unbound}, and the metastable
liquid region grows, in principle, for ever. Typical trajectories are shown in
Fig.\ \ref{trajectories}.

\begin{figure}
  \epsfxsize=8cm
  \begin{center}
  \leavevmode\epsffile{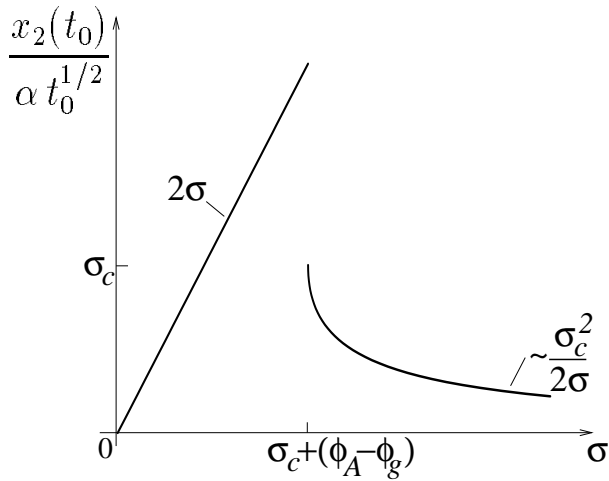}
  \caption{Size of the stable solid droplet $x_{2}$ at fixed  time
   $t_{0}$ ($t_{0}\rightarrow\infty$), against supersaturation
   $\sigma$; $\alpha^{-1} = (\phi_s-\phi_g)(\pi/D_g)^{1/2}$. For
   $\sigma'<\sigma_c$, the interfaces curvature-recombine at early
   times, and conventional growth takes place, at a rate proportional to
   $\sigma$. For $\sigma'>\sigma_c$, growth of the stable phase is increasingly
   suppressed. (Note that $\sigma_{c}\rightarrow 0$ at the triple point.)}
  \label{transition}
  \end{center}
  \vspace{-16pt}
\end{figure}

  In the unbound case, to reach the solid, material must diffuse through the
liquid phase (down a gradient which becomes flat at the triple point
$\sigma_c\to 0$); the growth rate of the solid region is thereby suppressed, by
a factor which we find to be of order $(\sigma_{c}/\sigma')^2$, compared to the
case of a curvature-bound interface \cite{Evans97b}. In this way, solid growth
from a supersaturated gas can be strongly arrested by a weakly metastable liquid
phase. Fig.\ \ref{transition} shows the size of the solid region $x_{2}$ at a
given late time $t_{0}$, as a function of supersaturation $\sigma$. For
$\sigma'>\sigma_{c}$, the solution of Eq.\ (\ref{x2}) is shown. For
$\sigma'<\sigma_{c}$, the metastable region collapses at some early time, after
which $j_{A}\approx\sigma\sqrt{D_{g}/\pi t}$.

  This scenario of diffusive unbinding of interfaces and suppression of crystal
growth relies on the initial formation of curvature-unbound interfaces during
the nucleation stage.  Whereas the critical
supersaturation $\sigma_{c}$ for diffusive unbinding is independent of
initial conditions, that for curvature-unbinding defines
a region, in the space of initial configurations, with a nontrivial boundary. In
our (non-exhaustive) numerical study of evolution of a step function,
curvature-unbinding appeared to occur whenever $\sigma'>\sigma_c$; if so, for
these initial states at least, the unbinding is diffusively controlled.

  Real systems are not infinite and hence the ambient supersaturation must
eventually be exhausted. If $L_{\mbox{\small nuc}}$ is the typical
inter-nuclear spacing, then the time to deplete the ambient supersaturation,
$t_{\mbox{\small dep}}$ (by growth of {\em either} solid {\em or} metastable
liquid) is of order $L_{\mbox{\small nuc}}^2/D_{g}$. The subsequent time-scale
for metastable regions to collapse and be replaced by crystals is of order
$(\sigma'/\sigma_{c})^2 t_{\mbox{\small dep}}$. This may explain, for example,
the suppression of rapid crystallization observed beyond a sharp boundary
\cite{footprime} in colloid-polymer phase diagrams \cite{Poon95}. As
$\sigma'$ approaches $\sigma_{c}$ from below, the lifetime of the metastable
phase diverges roughly as $85^{(1-\sigma'/\sigma_{c})^{-1/2}}$, consistent
with a {\em sharp} onset of arrested behavior.

  Notice finally that metastable phases are routinely ignored in ``diffusion
path analysis" (DPA), a method for predicting the evolution of
interfacial positions in the diffusive (long-time) regime by looking
solely at the {\em equilibrium} phase diagram. DPA has been widely
applied to systems undergoing {\em dissolution} \cite{Raney87}. We find
that, under dissolution conditions ($j_{A}<0$), {\em the g/s interface is always
diffusively bound} \cite{Evans97b}. This may account in part for the success of
DPA in that context.

  In summary we have shown that a metastable phase of intermediate density can
radically influence the dynamics of phase ordering in systems with a
single conserved order parameter. Such an intermediate can kinetically arrest
the formation of a dense phase. A fuller treatment of these phenomena will of
course require a three-dimensional approach (involving explicit treatment of
contact lines and interfacial curvature). It may also require inclusion of
additional order parameters to describe (say) local crystallinity (implicitly
treated above as rapidly relaxing). We leave
these issues for future work. However, we believe that our one-dimensional
treatment offers a strong physical picture of the arrest of domain growth in
conserved systems caused by metastable states, and that this will remain an
important part of any complete theory.

{\em Acknowledgements} This work was funded by EPSRC Grant No.\ GR/K56025.
W.C.K.P.\ thanks the Nuffield Foundation for a science fellowship.


\begin{thebibliography}{999}
\bibitem{nematic} Similar behaviour is seen in mixtures of colloidal
rods and low molecular weight polymer \protect\cite{vanBruggen96} where the
three phases are $I_{1}$, $I_{2}$ (both isotropic) and $N$ (nematic).
\bibitem{footprime} We suggest the boundary is at $\sigma'=\sigma_{c}$
rather than $\sigma'=0$ (the metastable binodal) as previously suggested
\cite{Poon95}. If $\sigma_{c}$ is small, it is experimentally difficult
to distinguish between the two, especially as no accurate theory exists
for $\phi_{A}$.
\end{thebibliography}
\end{document}